\documentclass[3p, 12pt, onecolumn]{elsarticle}
\usepackage{graphicx,amscd,amsmath,amssymb,verbatim}
\usepackage{amsfonts,epsfig}
\usepackage{mathptmx}
\usepackage{setspace}
\usepackage{epsfig}
\usepackage{url}
\usepackage{multirow}
\usepackage{subfigure}
\usepackage{graphicx,color}
\usepackage{algorithm,caption}
\usepackage{algpseudocode}
\usepackage{float}
\usepackage[normalem]{ulem}

\pdfoutput=1

\begin{document}

\journal{Physica A}

\begin{frontmatter}

\title{A physics-based algorithm to perform predictions in football leagues}

\author{Eduardo Velasco Stock$^{1}$, Roberto da Silva$^{1}$, Henrique A.
Fernandes$^{2}$}

\address{1 - Instituto de F\'{i}sica, Universidade Federal do Rio Grande do Sul,
Porto Alegre Rio Grande do Sul, Brazil\\
2 - Unidade Acadêmica de Ciências Exatas, Universidade Federal de Jataí, Jataí, Goiás, Brazil
}


\begin{abstract}

In this work, we extended a stochastic model for football leagues based on the team's 
potential [R. da Silva et al. Comput. Phys. Commun. \textbf{184} 661--670 (2013)] for 
making predictions instead of only performing a successful characterization of the 
statistics on the punctuation of the real leagues. Our adaptation considers the advantage 
of playing at home when considering the potential of the home and away teams. The algorithm 
predicts the tournament's outcome by using the market value or/and the ongoing team's 
performance as initial conditions in the context of Monte Carlo simulations. We present 
and compare our results to the worldwide known SPI predictions performed by 
the \textquotedblleft FiveThirtyEight\textquotedblright\ project. The results show that 
the algorithm can deliver good predictions even with a few ingredients and in more 
complicated seasons like the 2020 editions where the matches were played without fans in the stadiums.

\end{abstract}

\end{frontmatter}

\section{Introduction}

\label{Section:Introduction}

Football, if on the one hand, seems frivolous for a part of the world
population and socially unequal looking at the amounts of money received for
some players in major championships compared with those in many country-side
cities in Latin America well as other undeveloped countries, on the other
hand, it has its attractions. For example, it is professionally performed in
more than 200 countries according to FIFA (Federation Internationale de
Football Association) and generates many employments.

In addition, we have recently observed the excellent side of the social
pressure against the creation of the European Super League
(https://www.theguardian.com/football/2021/apr/20/european-super-league-unravelling-as-manchester-city-and-chelsea-withdraw). After all, the monetary discrepancies only reflect an unequal society since football is just a product of that same society.

Called soccer in the USA, this sport, which is the world's most popular one,
has other economic and social importance. For example, many advances in
sports science, medicine, and nutrition have maximized the players'
performance. Translating these points into money is difficult, but we have a
rough idea of how they can impact people's lives, especially the poorest
ones. Not necessarily, for this reason, physicists have devoted some work
and time to describe statistics related to football (see, for example, Refs. 
\cite{Heuer2009,Skinner2009}) in order to understand the stochastic and
deterministic aspects of this exciting research area.

Until the 1980s, the football's scoring system was 2-1-0, standing for
points by the win, draw, and loss, respectively. This system, however, had
benefited many "cross-country" teams, i.e., teams that only play to draw, as
this becomes very advantageous in this championship modality.

To make the game more attractive and, in a way, more competitive, an
evolution has started since then. The scoring system 3-1-0 prevailed over
the previous one, as well as other improvements. For example, when some
player now kicks back the ball to the goalkeeper, he/she is prohibited from
catching it with his/her hands. This simple rule increased the time of the
ball in motion in matches. Such rules are constantly updated, and all
improvements (or throwbacks), as well as new rules, can always be found in
the webpage related to "laws of the game" in the international football
association board (IFAB) \footnote{%
https://www.theifab.com/history/ifab}.

The magic point of football, which also occurs in many other sport
modalities, is that the best team does not always win the match, and in some
championships, "dark horses," a term used to describe a little known
candidate or competitor, who unexpectedly wins or succeeds, are ubiquitous.

In Brazil, this "upset victory" is called a "zebra," name created by Gentil
Cardoso, a Portuguesa iconic football coaching who told a reporter during an
interview about the possibility of his team beating Vasco da Gama in a match
valid for the regional championship in Rio de Janeiro, Brazil. Cardoso, at
the time, used the word "zebra " to refer to an animal not included in a
popular (but prohibited) gambling in Brazil, which we translate as "animal
game " in a free translation.

The so-called "zebras " are more frequent than one could imagine \cite%
{Zolkernevic2013,DASILVA2013661} and many times they break the deterministic
characteristics of championships, making the favorite teams not always the
champions \footnote{%
The reference \cite{Zolkernevic2013} is related to a report about football
organized by I. Zolkernevic highlighting the contribution in this area of
some Brazilian authors that instead of us. It includes also comments about
the work of Ribeiro et al. (see also \cite{Ribeiro2010}). The direct link to
this report is:
https://revistapesquisa.fapesp.br/en/upset-victories-common-brazilian-soccer-championship%
}. Commonly, the champion has a good cast of players, which, in turn,
depends, for instance, on the financial power of the team (although not
always). In balanced championships like those in Brazil, where many teams
have already become champions, the symbiosis of the team's players, combined
with a good coach, can even change history. Based on these assumptions, in
this paper, we intend to answer the following question: Are we able to
predict champions and relegations in a football championship using
mathematical models?

To shed light on this question, we first consider that each team can beat
its opponent, which, in turn, is translated as the probability of victory.
This probability depends on amounts defined as the team's potentials. These
potentials consider several factors, for example, the economic power of the
team and its history of successes along with the championship. This study
also takes into consideration a previous model that successfully described
the statistics of team scores in different double round-robin system (DRRS)
championships \cite{DASILVA2013661,PhysRevE.88.022136,DASILVA201456}, i.e.,
the championships in which all teams play each other twice, in turn, and
return.

This paper proposes a model that makes predictions of the champions, the
four best teams (G4), and the four worst teams (Z4) in a championship. For
this purpose, we adopted two different parameters, the market value, and the
teams' performance, as shown in Section \ref{Sec:Model}. In Section \ref%
{Sec:Results}, we divide our results into two different parts: the first one
corresponds to an exploration of the tuning/calibration of some parameters
with special attention to the influence of market value and memory effects
on modeling and, in the second part of the results, we take into account the
optimal parameters obtained previously to test our model presenting the
predictions for the Brazilian Championship A Series and comparing them with
some specialized websites. Finally, some conclusions are presented in
Section \ref{Sec:Conclusions}.

\section{Model and optimization}

\label{Sec:Model}

Our model is based on the agent-based model \cite%
{DASILVA2013661,PhysRevE.88.022136,DASILVA201456} that considers a system of 
$N$ teams playing against each other according to DRRS \footnote{%
https://en.wikipedia.org/wiki/Round-robin\_tournament}. Here, it is
noteworthy that although the goal of this study is to make predictions about
the top Brazilian professional league for men's football clubs called
Brazilian Championship A Series (commonly referred as \textquotedblleft
Brasileir\~{a}o\textquotedblright), our approach is generic and then, can by
applied to other leagues as well.

After a given season starts, if team $i$ plays as host (\textit{home team})
against team $j$ as visitor (\textit{away team}) at the $k$-th round, and
based on the fact that the number of goals in a match follows a Poisson
distribution \cite{Skinner2009,MAHER,Heuer2010}, the probability of the
match to result in a draw can be written as \cite{DASILVA2013661}: 
\begin{equation}
\begin{array}{ccc}
r_{draw}^{(i,j)}(k) & = & \Pr \left[ (g_{i}=g_{j})|(\phi _{k}^{(i)},\psi
_{k}^{(j)})\right] \\ 
&  &  \\ 
& = & \sum_{g=0}^{\infty }\frac{\left( \phi _{k}^{(i)}\psi _{k}^{(j)}\right)
^{g}}{{g!}^{2}}e^{-\left( \phi _{k}^{(i)}\psi _{k}^{(j)}\right) } \\ 
&  &  \\ 
& = & e^{-\left( \phi _{k}^{(i)}\psi _{k}^{(j)}\right) }I_{0}\left( 2\sqrt{%
\phi _{k}^{(i)}\psi _{k}^{(j)}}\right)%
\end{array}
\label{prob_draw}
\end{equation}%
where $g$ is the number of goals scored by each team, $\phi $ and $\psi $
correspond to the host's and the visitor's potentials respectively, and 
\begin{equation*}
I_{\nu }(z)=\left( \frac{1}{2}\right) ^{\nu }\sum_{n=0}^{\infty }\frac{%
\left( \frac{1}{4}z^{2}\right) ^{n}}{n!\Gamma (\nu +n+1)}
\end{equation*}%
is the modified Bessel function of the first kind. Our reason to define a
home team potential ($\phi $) and an away team potential ($\psi $), resides
on the well known fact that teams that are hosting the match have
statistical advantage over their visitors \cite{pollard1986,GOUMAS2017321}
for a number of reasons such as travelling distances, crowd size, number of
time zones crossed by the visiting team, altitude of the home stadium etc.

The probability that team $i$ has of winning the match against team $j$ at $%
q $-th round considers the product of the complement of Eq. \ref{prob_draw}
and a factor that weights the potentials of the teams involved such as
follows: 
\begin{equation}
\Pr (g_{i}>g_{j},q)=\left[ 1-r_{draw}^{(i,j)}(q)\right] \cdot \frac{\phi
_{q}^{(i)}}{\phi _{q}^{(i)}+\psi _{q}^{(j)}}\text{.}
\label{Eq:prob_home_win}
\end{equation}
The chances of the team $j$ to win the match is defined by the complement: 
\begin{equation}
\begin{array}{lll}
\Pr (g_{i}<g_{j},q) & = & 1-\Pr (g_{i}>g_{j},q)-r_{draw}^{(i,j)}(q) \\ 
&  &  \\ 
& = & \left[ 1-r_{draw}^{(i,j)}(q)\right] \cdot \frac{\psi _{q}^{(j)}}{\phi
_{q}^{(i)}+\psi _{q}^{(j)}}\text{.}%
\end{array}
\label{Eq:prob_away_win}
\end{equation}

The team's potential is updated after each match. Therefore, we must
consider the final result of the match, i.e., whether the team won, lost or
ended in a draw. With this in mind, we defined that the winner and the loser
teams will have their potentials changed respectively by the quantities ${%
\Delta \phi }_{q}$ and ${\Delta \psi }_{q}$. By considering that the home
team $i$ plays against the visiting team $j$, we have 
\begin{equation*}
{\Delta \phi }_{q}^{(i)}=\left\{ 
\begin{array}{cccc}
-{\Delta \psi }_{q}^{(j)}=3 &  & \text{if } & g_{i}^{(q)}>g_{j}^{(q)}, \\ 
&  &  &  \\ 
{\Delta \psi }_{q}^{(j)}=1 &  & \text{if } & g_{i}^{(q)}=g_{j}^{(q)}, \\ 
&  &  &  \\ 
-{\Delta \psi }_{q}^{(j)}=-3 &  & \text{if } & g_{i}^{(q)}<g_{j}^{(q)},%
\end{array}%
\right.
\end{equation*}%
yielding ${\Delta \psi }_{q}^{(i)}={\Delta \phi }_{q}^{(j)}=0$ in all
conditions.

Before any season starts, the market value \footnote[1]{%
https://www.transfermarkt.com/} of each team is supposed to be one of the
main factors to reflect its potential performance, which has influence on
its chances of becoming the future champion of the league. After the start
of a season, the real performance of the teams comes to light as new and
important information that must be taken into account in predictive
algorithms. Thus, we define the both potentials of a team $i$ at the $q$-th
round as: 
\begin{equation}
\begin{array}{ccc}
\phi _{q}^{(i)} & = & \alpha M^{(i)}+\beta P_{k_{0}|\phi }^{(i)}+\gamma
\sum\limits_{j=k_{0}}^{q-1}\Delta \phi _{j}^{(i)} \\ 
&  &  \\ 
{\psi }_{q}^{(i)} & = & \alpha M^{(i)}+\beta P_{k_{0}|{\psi }}^{(i)}+\gamma
\sum\limits_{j=k_{0}}^{q-1}\Delta {\psi }_{j}^{(i)},%
\end{array}
\label{Eq:potentials_evolutions}
\end{equation}
where $\phi _{k_{0}}^{(i)}=\alpha M^{(i)}+\beta P_{k_{0}|\phi }^{(i)}$ and ${%
\psi }_{k_{0}}^{(i)}=\alpha M^{(i)}+\beta P_{k_{0}|{\psi }}^{(i)}$ are the
initial potentials, $M^{(i)}$ is the market value of team $i$, and
therefore, $\alpha $ is the market coefficient. Similarly, we also use the
scored points to perform predictions, thus $P_{k_{0}}^{(i)}$ is the
cumulative scored points in the first $k_{0}$ rounds and $\beta $ is a
coefficient related to the performance of the team. The quantities $%
P_{k_{0}|\phi }^{(i)}$ and $P_{k_{0}|{\psi }}^{(i)}$ are related to the
cumulative scored points as the team plays at home and when it plays as a
visitor, respectively, and of course, $P_{k_{0}}^{(i)}=P_{k_{0}|\phi
}^{(i)}+P_{k_{0}|{\psi }}^{(i)}$. Finally, we complete the contribution for
the potentials by including the sum of the successive increments/decrements
from $k_{0}$ up to the $q$-th round, i.e., $\sum\limits_{j=k_
{0}}^{q-1}\Delta \phi _{j}^{(i)}$ and $\sum\limits_{j=k_{0}}^{q-1}\Delta {%
\psi }_{j}^{(i)}$. Thus, $\gamma $ is a coefficient associated to this
quantity.

Now, it comes the problem: how to determine $\alpha $, $\beta $, and $\gamma 
$? Once these values are estimated, we are able to obtain $\phi _{q}^{(i)}$
and ${\psi }_{q}^{(i)}$ for arbitrary $q$-th round. To reach this goal, we
define $k$ such that a memory of $\Delta k=k-k_{0}$ works as a
\textquotedblleft learning interval\textquotedblright\ to determine the
optimal values $\alpha _{opt}$, $\beta _{opt}$, and $\gamma _{opt}$, which
maximizes relevant parameters that compare the results from simulation and
real matches obtained in this interval. Thus, let us explain this process
that basically works with three algorithms.

The main algorithm (Algorithm 1) defines the evolution of the potentials
described by Eq. (\ref{Eq:potentials_evolutions}) in a general way, as well
as the evolution of the scores.

\begin{algorithm}[H]
\caption*{\textbf{Algorithm 1: Main}}
\label{alg:main_alg}
\begin{algorithmic}[1]
\If {$\left(rand[0,1] < r_{draw} \right)$}
    \State $P^{(i)}=P^{(i)}+1$ and $P^{(j)}=P^{(j)}+1$
\Else
    \If{$\left( rand[0,1] < \left( \frac{\phi^{(i)}}{\phi^{(i)}+\psi^{(j)}}\right)\right)$}
        \State $P^{(i)}=P^{(i)}+3$; $\phi^{(i)}=\phi^{(i)}+\gamma*\Delta \phi^{(i)}$; $\psi^{(j)}=\psi^{(j)}+\gamma*\Delta \psi^{(j)}$;
    \Else
        \State $P^{(j)}=P^{(j)}+3$; $\psi^{(j)}=\psi^{(j)}+\gamma*\Delta \psi^{(j)}$; $\phi^{(i)}=\phi^{(i)}+\gamma*\Delta \phi^{(i)}$;
    \EndIf
\EndIf
\end{algorithmic}
\end{algorithm}

First, it is necessary to perform the optimization. For that, we have
alternatives based on some metrics. We define our first metrics denoted by 
\textit{matching hits}: 
\begin{equation}
\xi _{q}=\frac{2}{N}\sum\limits_{l=1}^{N/2}\delta (\varphi
_{q}^{(l)}|_{S},\varphi _{q}^{(l)}|_{R})  \label{Eq:matchhits}
\end{equation}%
where $\varphi _{q}^{(l)}|_{S}$ and $\varphi _{q}^{(l)}|_{R}$ are the
results of the $l$-th match in the $q$-th round of the simulated (S) and the
real world (R), respectively, and $\delta (x,y)$ is the well-known Kronecker
symbol. Here, one considers only the result and the exact number of goals is
not taken into account. Thus, in the tuning algorithm (Algorithm 2), we
perform the optimization of our model in order to obtain the values $\alpha
_{opt}$, $\beta _{opt}$, and $\gamma _{opt}$ as presented below.

\begin{algorithm}[H]
\caption*{\textbf{Algorithm 2: Tuning}}
\label{alg:tun_alg}
\begin{algorithmic}[1]
\For {$\alpha=0$ \textbf{to} $\alpha_{max}$ \textbf{step} $\Delta \alpha$}
  \For {$\beta=0$ \textbf{to} $\beta_{max}$ \textbf{step} $\Delta \beta$}
    \For {$\gamma=0$ \textbf{to} $\gamma_{max}$ \textbf{step} $\Delta \gamma$}
        \For {$i_{run}=1$ \textbf{to} $N_{run}$}
            \For {$i_{round}=k_{0}$ \textbf{to} $k$}
                \State run \textbf{Main Algorithm};
	        \EndFor
	    \EndFor
	    \If { $\xi > \xi_{max}$}
	        \State $\alpha_{opt}=\alpha$; $\beta_{opt}=\alpha$; $\gamma_{opt}=\gamma$; $\xi_{max}=\xi$;
	    \EndIf
        \EndFor
    \EndFor
\EndFor
\end{algorithmic}
\end{algorithm} 

The last algorithm, the Algorithm 3 shown in the following, is designed to
perform the final evolution of the simulation delivering the championship
ranking table by considering the values of $\alpha _{opt}$, $\beta _{opt}$,
and $\gamma _{opt}$ obtained previously.

\begin{algorithm}[H]
\caption*{\textbf{Algorithm 3: Final evolution}}
\label{alg:tun_alg}
\begin{algorithmic}[1]
\For {$i_{run}=1$ \textbf{to} $N_{run}$}
    \For {$i_{round}=k$ \textbf{to} $N_{last}$}
        \State run \textbf{Main Algorithm} using $\alpha_{opt}$, $\beta_{opt}$, and $\gamma_{opt}$;
    \EndFor
\EndFor
\end{algorithmic}
\end{algorithm}

The purpose of each of these algorithms can be summarized as follows

\begin{enumerate}
\item Algorithm 2 calls Algorithm 1 $N_{run}$ times. The simulation
considers the current championship, from $k_{0}$ up to the $k$-th round for
the variables $\alpha$, $\beta$, and $\gamma $, using as input $\phi
_{k_{0}}^{(i)}$, $\psi _{k_{0}}^{(i)}$, $P_{k_{0}|\phi}^{(i)}$, and $%
P_{k_{0}|{\psi }}^{(i)}$ for $i=1,...,N$, in order to obtain the average
value of $\xi _{q}$ for these $\frac{N}{2}\Delta k$ matches.

\item By performing three external loops, Algorithm 2 increments the values
of $\alpha$, $\beta$, and $\gamma$ by the amount $\Delta \alpha$, $\Delta
\beta$, and $\Delta \gamma$ and then, we repeat the step 1 for each set $%
\alpha$, $\beta$, and $\gamma \in \lbrack 0,1]$ searching for $\alpha _{opt}$%
, $\beta _{opt}$, and $\gamma_{opt})$ that leaded to a greater average value
of $\xi _{q}$.

\item With $\alpha _{opt}$, $\beta _{opt}$, and $\gamma _{opt}$ in hand, one
runs $N_{run}$ times once again the championship from the round $k$ until
the last round ($N_{last}=2(N-1)$, where $N$ is the number of teams) in
order to obtain the average results of each team (how much times it becomes
the champion, its participation in $g_{4}$, and its participation in the $%
z_{4}$), according to Algorithm 3 (final evolution).
\end{enumerate}

In addition, we can use alternative metrics to match the hits defined in Eq.
(\ref{Eq:matchhits}). One interesting metric is based on two amounts: one
based on scored points and one based on variable ranking. The first one
measures the predictive capability of our algorithm to measure the
difference of the scored points in simulations (S) and those obtained of the
real world (R), which are given by $P_{k}|_{S}$ and $P_{k}|_{R}$ per team,
respectively, until a $q$-th round defined as 
\begin{equation}
\mu _{q}=\frac{1}{N}\sum\limits_{l=1}^{N}\eta \left(
P_{q}^{(l)}|_{S},P_{q}^{(l)}|_{R}\right) ,  \label{Eq:mu}
\end{equation}%
where 
\begin{equation*}
\eta (x,y)=\left\{ 
\begin{array}{ll}
1 & \text{if}\quad |x-y|\ \leq \epsilon \\ 
&  \\ 
0 & \text{otherwise}%
\end{array}%
\right.
\end{equation*}%
and $\epsilon $ is an error parameter. If $\epsilon =0$, we have what we
called the \textit{strong} predictor. On the other hand, $\epsilon =1$ is
called \textit{medium} predictor and $\epsilon =2$ means \textit{weak}
predictor. The ranking-based variable takes into account the difference
between the team's ranking position in the simulation and the real league
classification table as follows 
\begin{equation}
\nu _{q}=\frac{1}{N}\sum\limits_{l=1}^{N}\eta \left(
Q_{q}^{(l)}|_{S},Q_{q}^{(l)}|_{R}\right) ,  \label{Eq:nu}
\end{equation}%
where $Q_{q}^{(l)}$ denotes the ranking position of the team $l$ at the $q$%
-th round.

The search for the values of $\alpha $, $\beta $, and $\gamma $ that
maximizes $\xi _{q}$ in Eq. (\ref{Eq:matchhits}) is simply changed by 
\begin{equation*}
\xi _{q}=\mu _{q}+\nu _{q}
\end{equation*}%
which is denoted as \textit{score-ranking metric}. This metric mixes the
influence of the rank and score of the teams simultaneously and, for this
reason, we performed the predictions using this metric instead of the
matching hits. In this paper, we decided to use $\epsilon =0$, i.e., the
strongest criterion for our approach.

\section{Results}

\label{Sec:Results}

\begin{figure}[]
\begin{center}
\includegraphics[width=0.5%
\columnwidth]{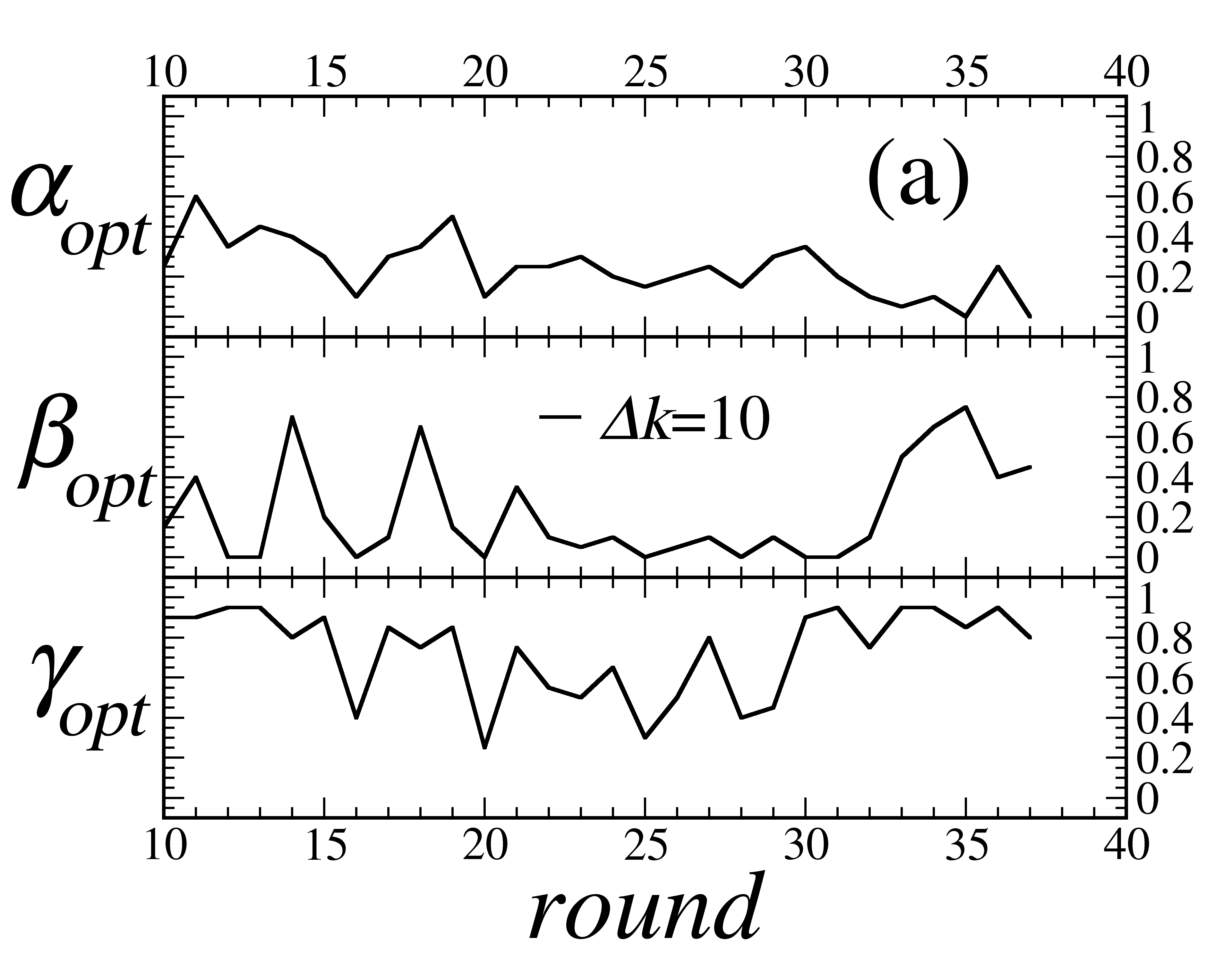}%
\includegraphics[width=0.5%
\columnwidth]{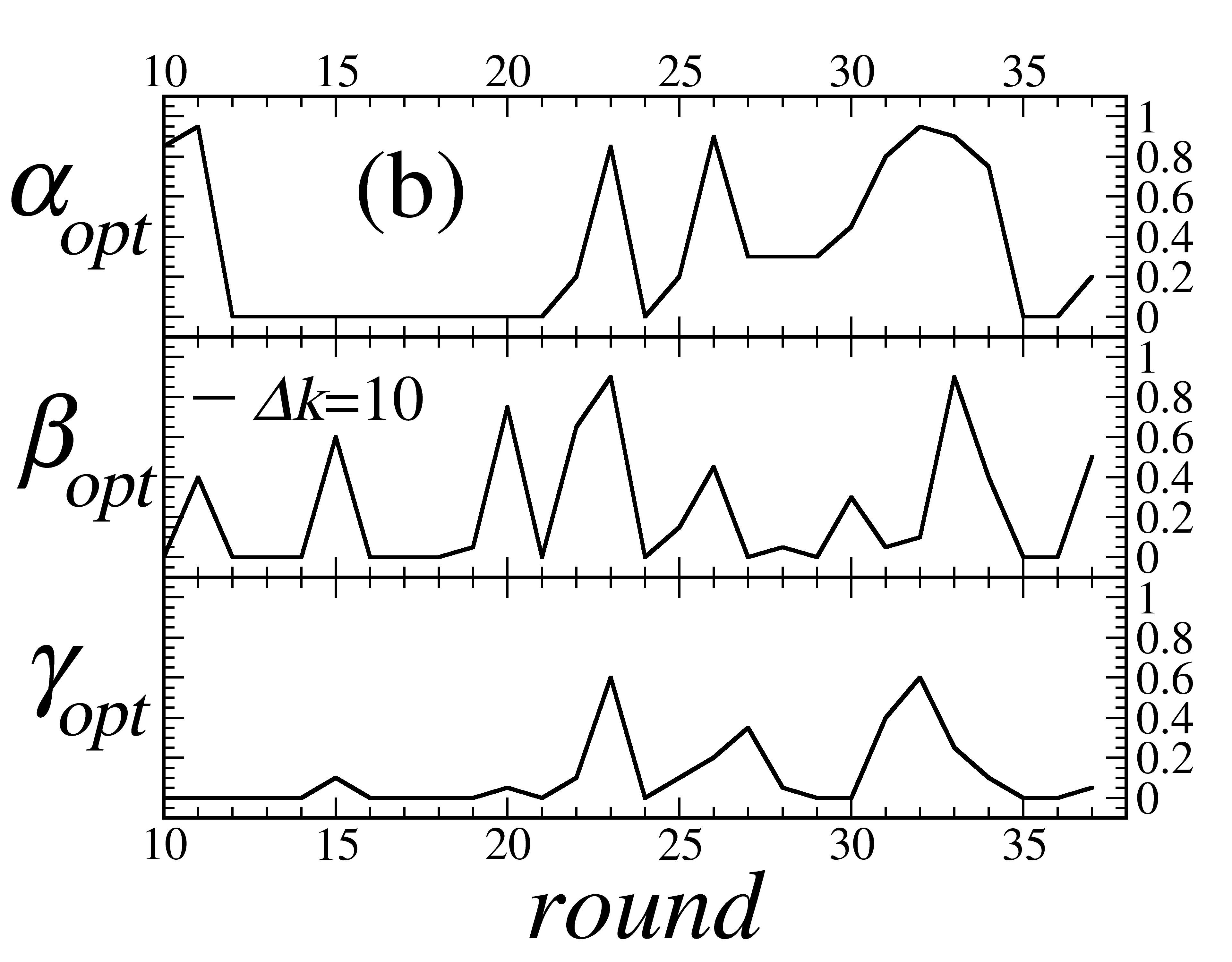}
\end{center}
\caption{Evolution of the parameters $\protect\alpha _{opt}$, $\protect\beta%
_{opt}$, and $\protect\gamma _{opt}$ per round for the 2020 edition of the
Brazilian Championship A Series. (a) corresponds to the case where we used
the matching hits criterion and (b) corresponds to the case where the
criterion was the maximization of the $\protect\mu +\protect\nu $.}
\label{Fig1}
\end{figure}

First, we explore how the Algorithm 2 works and in this preparatory part, we
will show how the parameters behave. We consider $N_{run}=100$ and $\Delta
\alpha =\Delta \beta =\Delta \gamma =0.05$. In Fig. \ref{Fig1}, we present
the evolution of these parameters for the 2020 edition of the Brazilian
Championship A Series considering the two criteria for the optimization:
highest average value of matching hits as shown in Fig. \ref{Fig1} (a) and
highest average value of the $\mu +\nu $ (score-ranking metric) presented in
Fig. \ref{Fig1} (b).

In Fig. \ref{Fig1} (a), we observe that with the matching hits criterion,
the coefficients seem to fluctuate slightly less than when considering the
score-ranking metric (Figure \ref{Fig1} (b)), mainly when one observes the
evolution of $\alpha $ which is related to the market value of the team. As
can be seen, we considered $\Delta k=10$ in this study. An interesting
analysis can be performed if we fix $\beta $ and $\gamma $ focusing only in
behavior of $\alpha $ for different rounds. In this case, we find by the
optimal value of $\alpha $ and consider $\beta =\gamma =1$. This analysis
helps us to understand how much of the market value must be added to perform
good predictions.

\begin{figure}[]
\begin{center}
\includegraphics[width=0.5%
\columnwidth]{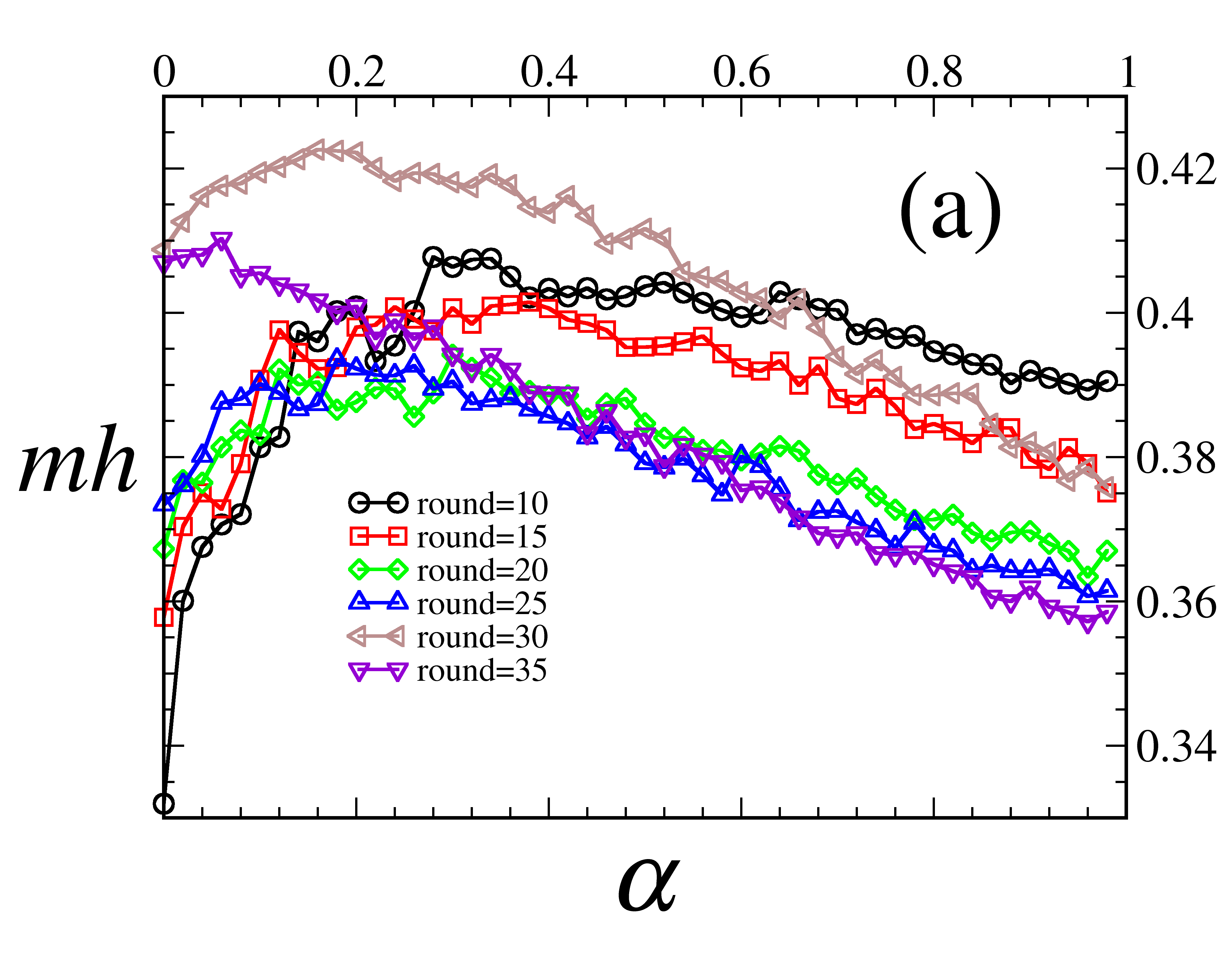}%
\includegraphics[width=0.5%
\columnwidth]{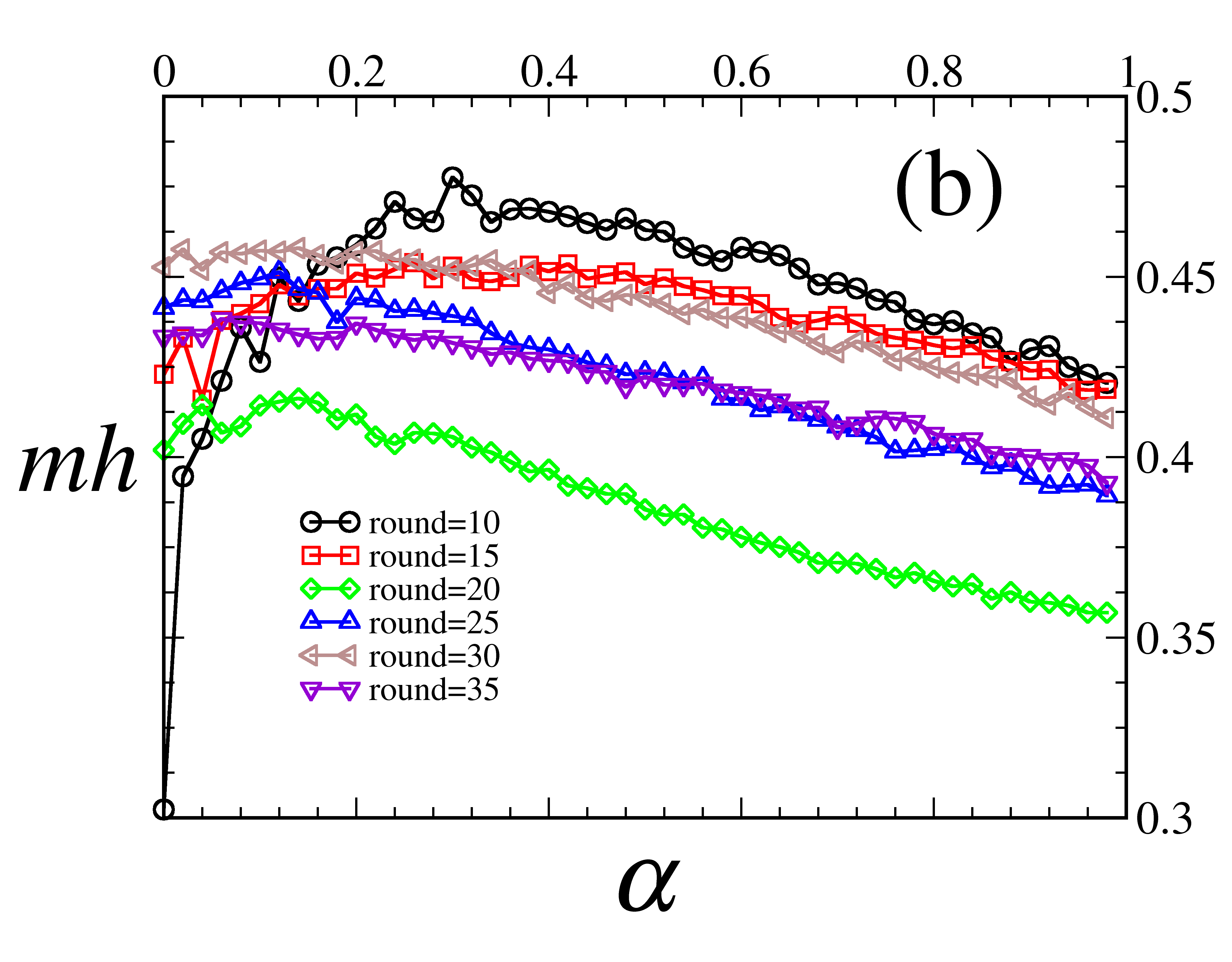} %
\includegraphics[width=0.5%
\columnwidth]{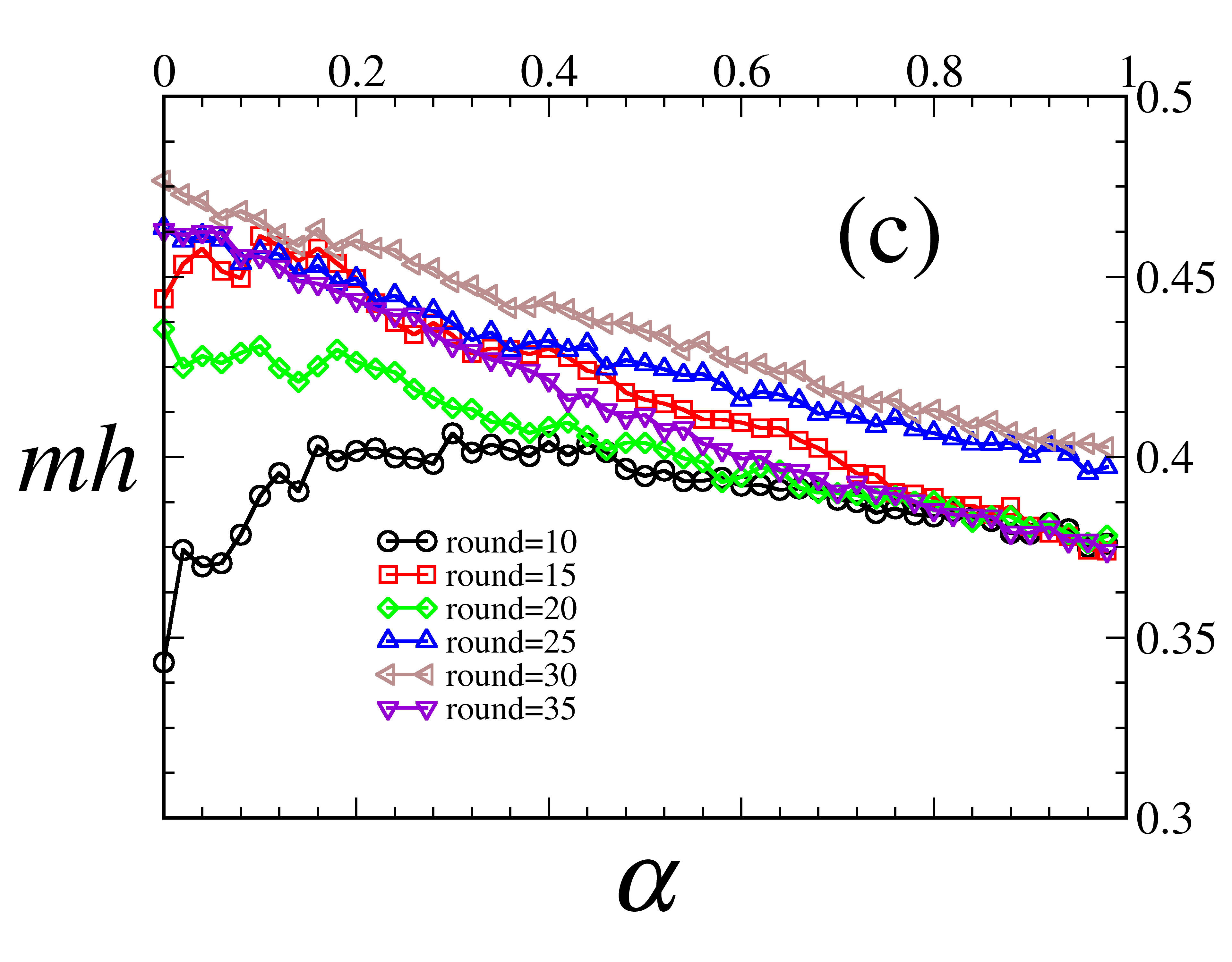}%
\includegraphics[width=0.5%
\columnwidth]{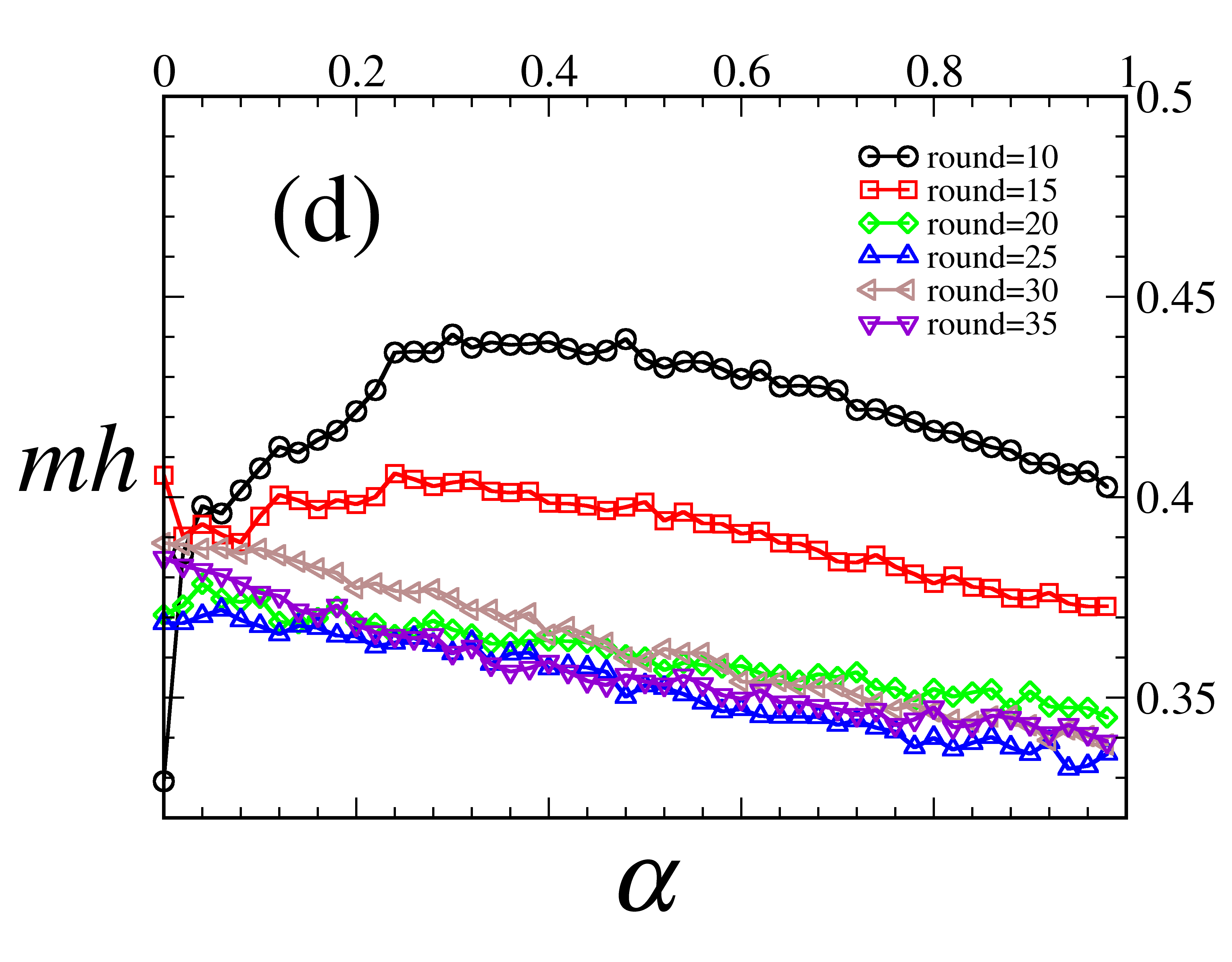}
\end{center}
\caption{Matching hits versus $\protect\alpha$ for different rounds: $k=10$, 
$15 $, $20$, $25$, $30,$ and $35$. Plots (a), (b), (c), and (d), shows the
results for the seasons 2020, 2019, 2018, and 2017, respectively. In all
situations one used $\Delta k=10$, $N_{run}=1000$, and $\protect\beta =%
\protect\gamma =1$.}
\label{Fig: Match_hits_versus_alfa}
\end{figure}

Figure \ref{Fig: Match_hits_versus_alfa} shows the behavior of matching
heats (mh) parameter as function of $\alpha $ for different rounds and
several seasons of the considered Brazilian Championship A Series. We fixed $%
\beta =\gamma =1$ to perform these plots.

Particularly for season 2020 shown in Fig. \ref{Fig: Match_hits_versus_alfa}%
, an \textquotedblleft almost\textquotedblright\ monotonic decay is observed
only for the largest $k$, $k=35$. For the other cases with $k\leq 30$, one
observes an initial increase and a subsequent decrease of this quantity as
function of $\alpha $, showing that $\alpha_{opt}$ corresponds to the peak
of these curves. However, for the other seasons, one observes that even for
intermediate values of $k$, this \textquotedblleft almost\textquotedblright\
monotonic decay behavior is already observed indicating less influence of
the market value on the matching hits.

\begin{figure}[]
\begin{center}
\includegraphics[width=0.5\columnwidth]{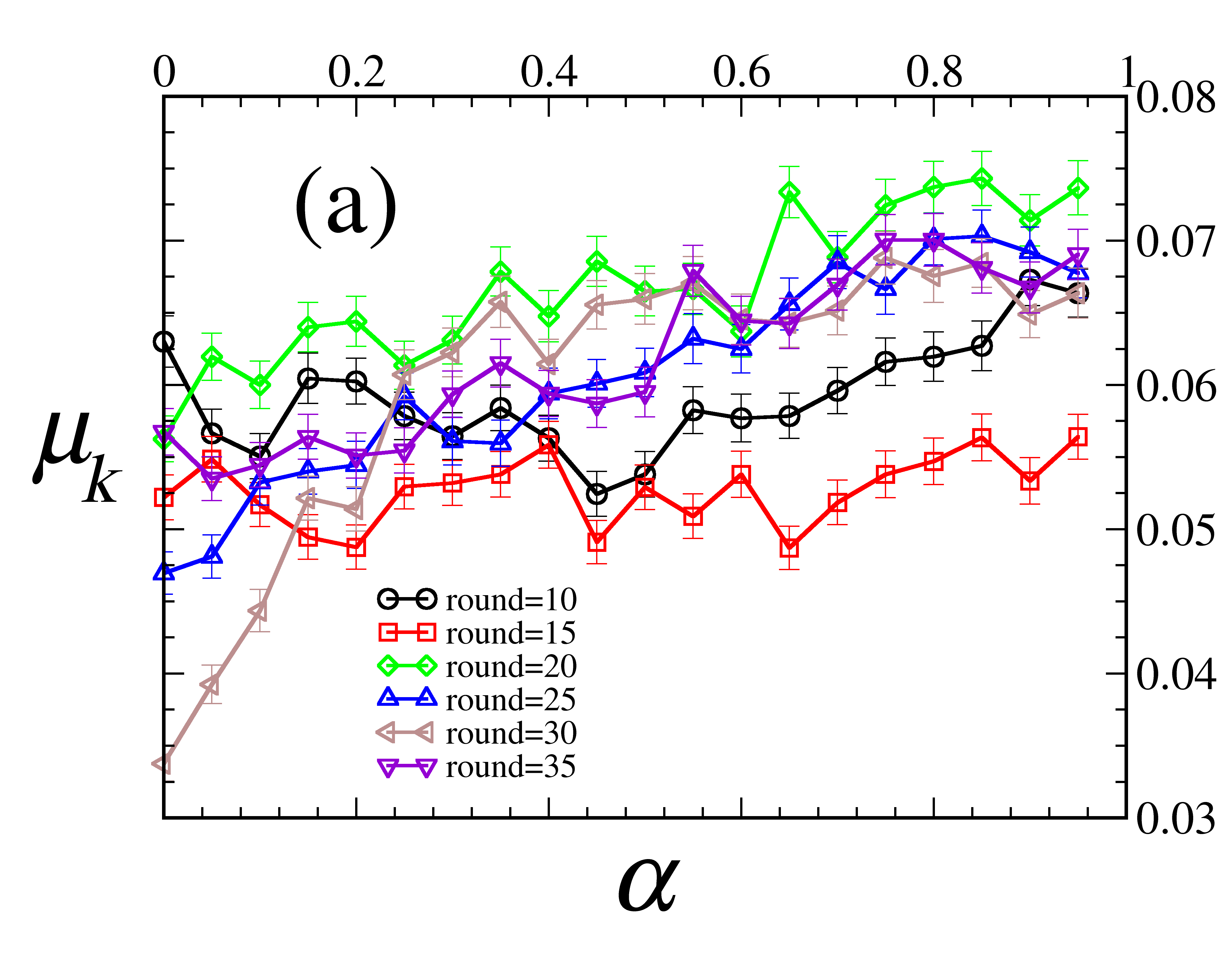}%
\includegraphics[width=0.5\columnwidth]{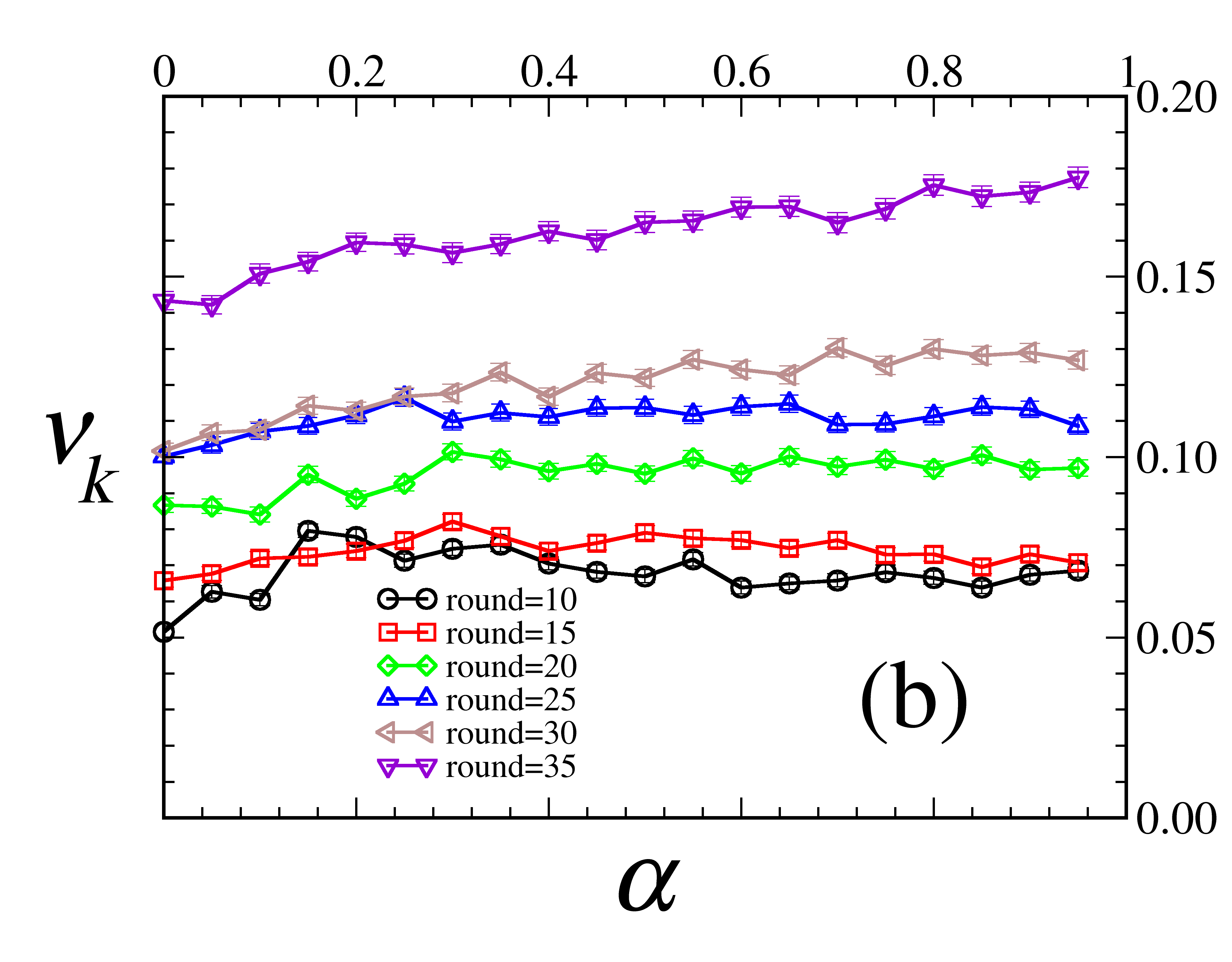}
\end{center}
\caption{Parameters of score $\protect\mu_{k}$ and ranking $\protect\nu_{k}$
as function of $\protect\alpha $ for different rounds $k$.}
\label{Fig:mu_and_nu_as_function_of_alfa}
\end{figure}

Differently from the matching hits, the parameters to compose the
score-ranking metric $\mu $ and $\nu $ present slow increasing tendency on $%
\alpha $ as can be observed in Fig. \ref{Fig:mu_and_nu_as_function_of_alfa}
(a) and (b), respectively.

Without loss of generality, for saving space, in Fig. \ref%
{Fig:mu_and_nu_as_function_of_alfa}, we only present the results for 2020
season since it is the most debatable season. Although not presented here,
the other seasons have similar behavior. Since we show the way to obtain the 
$\alpha_{opt}$, it is interesting to present a typical plot of $\alpha_{opt}$
as function of round ($k$). Figure \ref%
{Fig:Evolution_of_alfa_fixing_beta_gamma} shows our results by using the two
considered criteria: matching hits and score-ranking metric. Fig \ref%
{Fig:Evolution_of_alfa_fixing_beta_gamma} (a) presents our estimates for the
matching hits criterion and Fig. \ref%
{Fig:Evolution_of_alfa_fixing_beta_gamma} (b) shows the results for the
score-ranking one.

\begin{figure}[]
\begin{center}
\includegraphics[width=0.5%
\columnwidth]{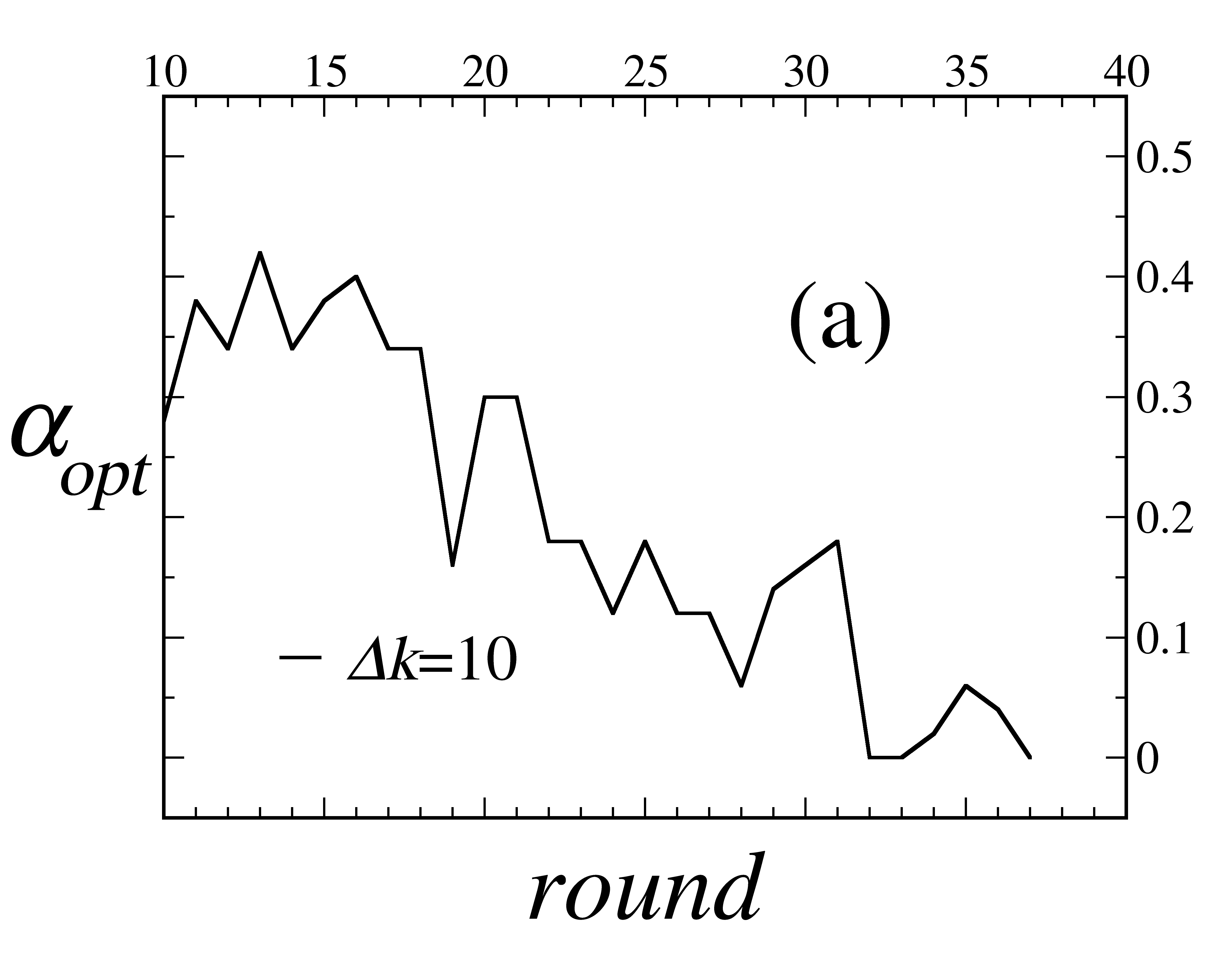}%
\includegraphics[width=0.5%
\columnwidth]{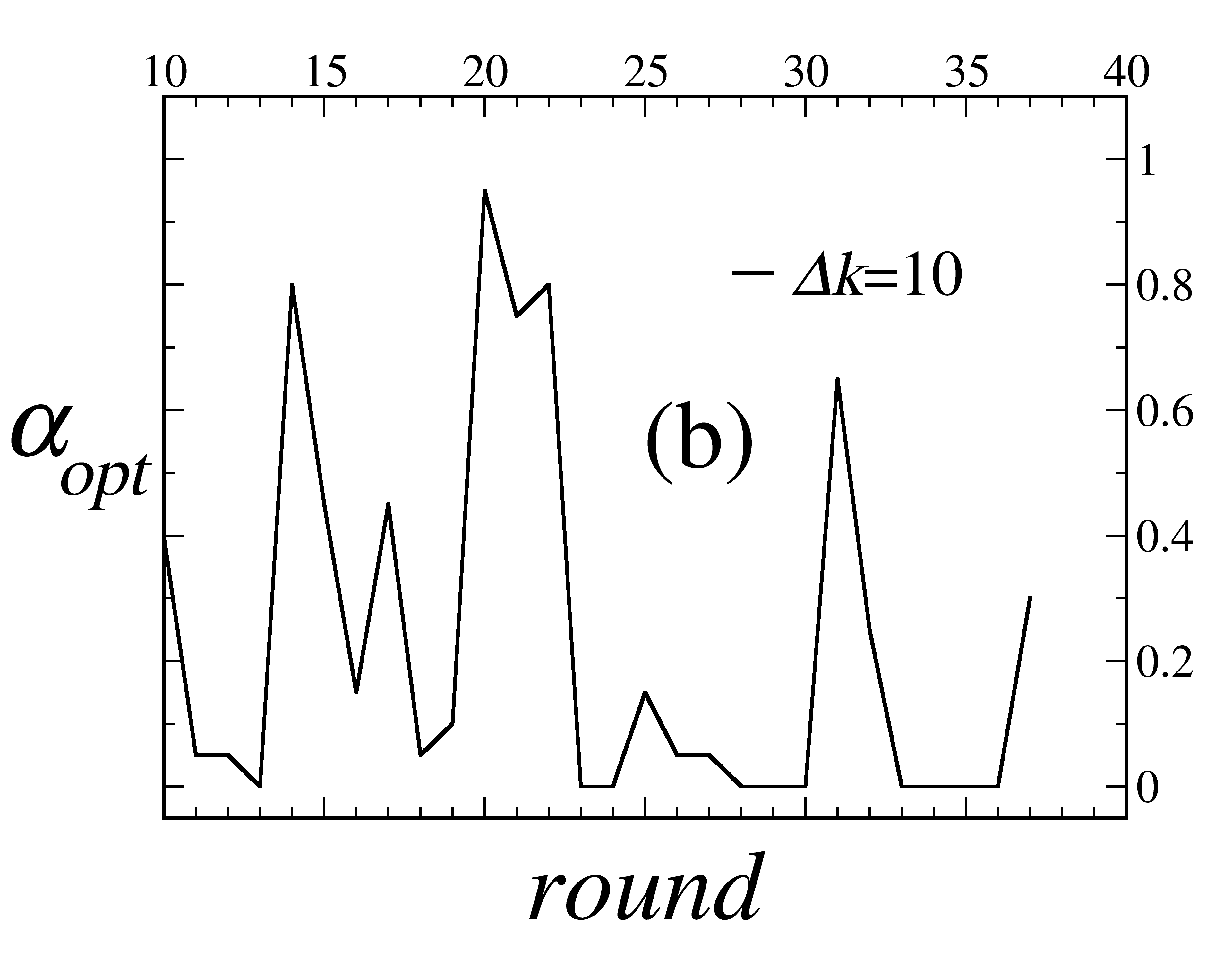}
\end{center}
\caption{Evolution of the parameter $\protect\alpha_{opt}$ per round keeping
fixed $\protect\beta=\protect\gamma=1$, for the 2020 edition of the
Brazilian Championship A Series. (a) corresponds the case where we used
matching hits criterion, and (b) is the case where the criteria was the
maximization of the $\protect\mu+\protect\nu$. }
\label{Fig:Evolution_of_alfa_fixing_beta_gamma}
\end{figure}
One can observe that $\alpha _{opt}$ has a tendency to decrease as $k$
increases by using the matching hits criterion. On the other hand, $%
\alpha_{opt}$ varies strongly with the score-ranking criterion. This
scenario must be reflected in our study.

\begin{figure}[]
\begin{center}
\includegraphics[width=1.0%
\columnwidth]{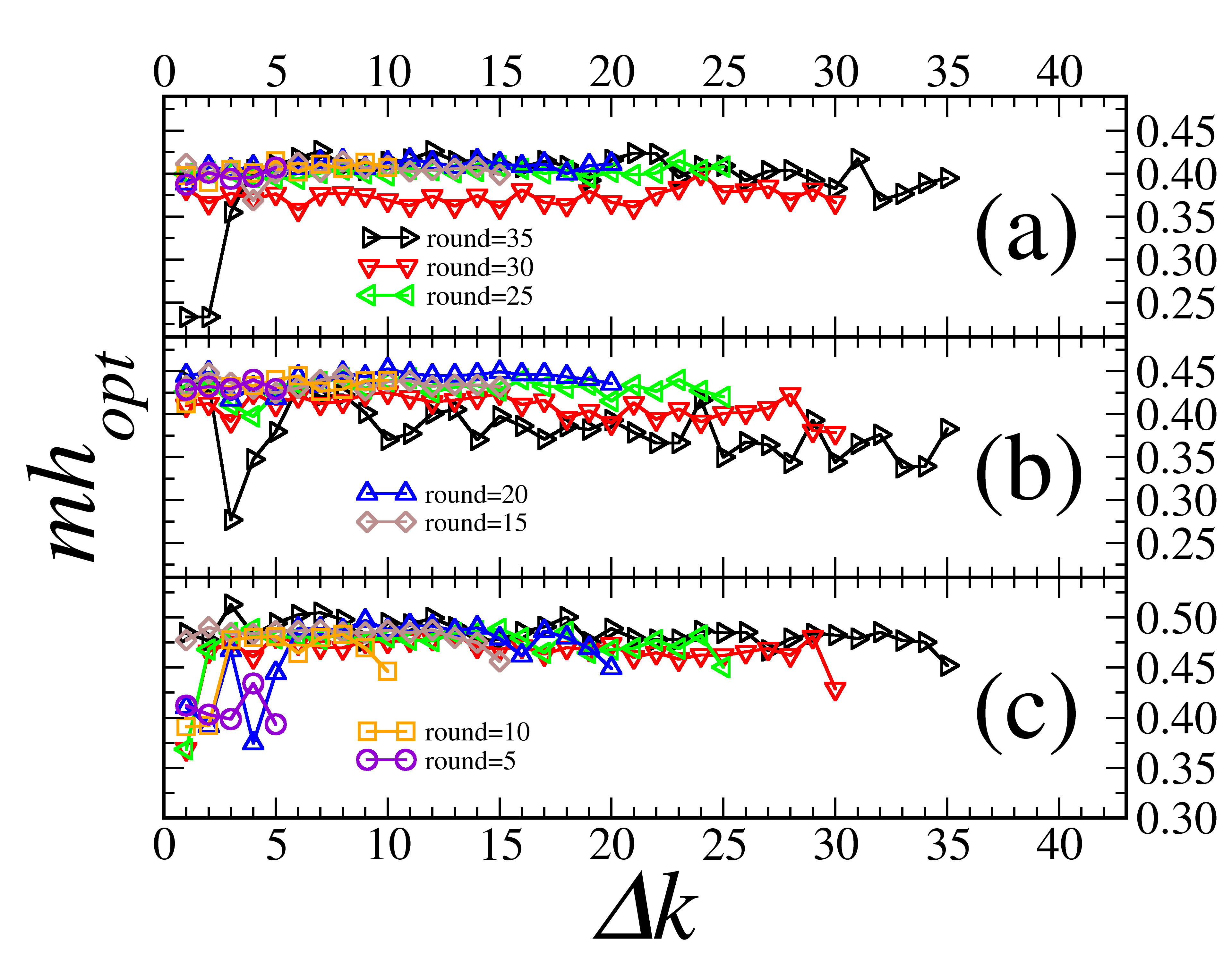}
\end{center}
\caption{$mh_{opt}$, which denotes the average matching hits obtained with $%
\protect\alpha_{opt}$, as function of $\Delta k$ for 2018 (a), 2019 (b), and
2020 (c) editions of the championship}
\label{Fig:Optimal_match_hits}
\end{figure}

Once we understand how to obtain the optimal parameters, we are able to
continue the study of the proposed model in order to make predictions of the
championship results. Some questions should be raised in relation to the
memory $\Delta k$ and, to answer them we will consider only the matching
hits metric. In Fig. \ref{Fig:Optimal_match_hits}, we show a plot of $%
mh_{opt}$, denoting the average matching hits calculated for the $%
\alpha_{opt}$, as function of $\Delta k$ for three seasons of the Brazilian
Championship A series: 2018, 2019, and 2020, and keeping fixed $%
\beta=\gamma=1$.

From this figure, we can observe that $\alpha_{opt}$ presents small
fluctuations around a mean value for $\Delta k\geq 10$ and, for this reason,
we will use this value until the end of this paper. But the question is
whether such optimal values can bring good predictions for the
championships. As we will show below, the answer is yes and the optimal
parameters used as input in Algorithm 3 are capable to deliver good
predictions to the results of the championships.

To study the predictive abilities of our model, we feed Algorithm 3 with
data of 2017, 2018, 2019 and 2020 seasons of the Brazilian Championship A
Series in order to be able to compare our simulations outcome with the final
results of the seasons. However, the 2020 season will deserve more attention
because it was a particularly complex season as it was decided only in the
last round and held during the Covid-19 pandemic.

\begin{figure*}[h]
\begin{center}
\includegraphics[width=%
\textwidth]{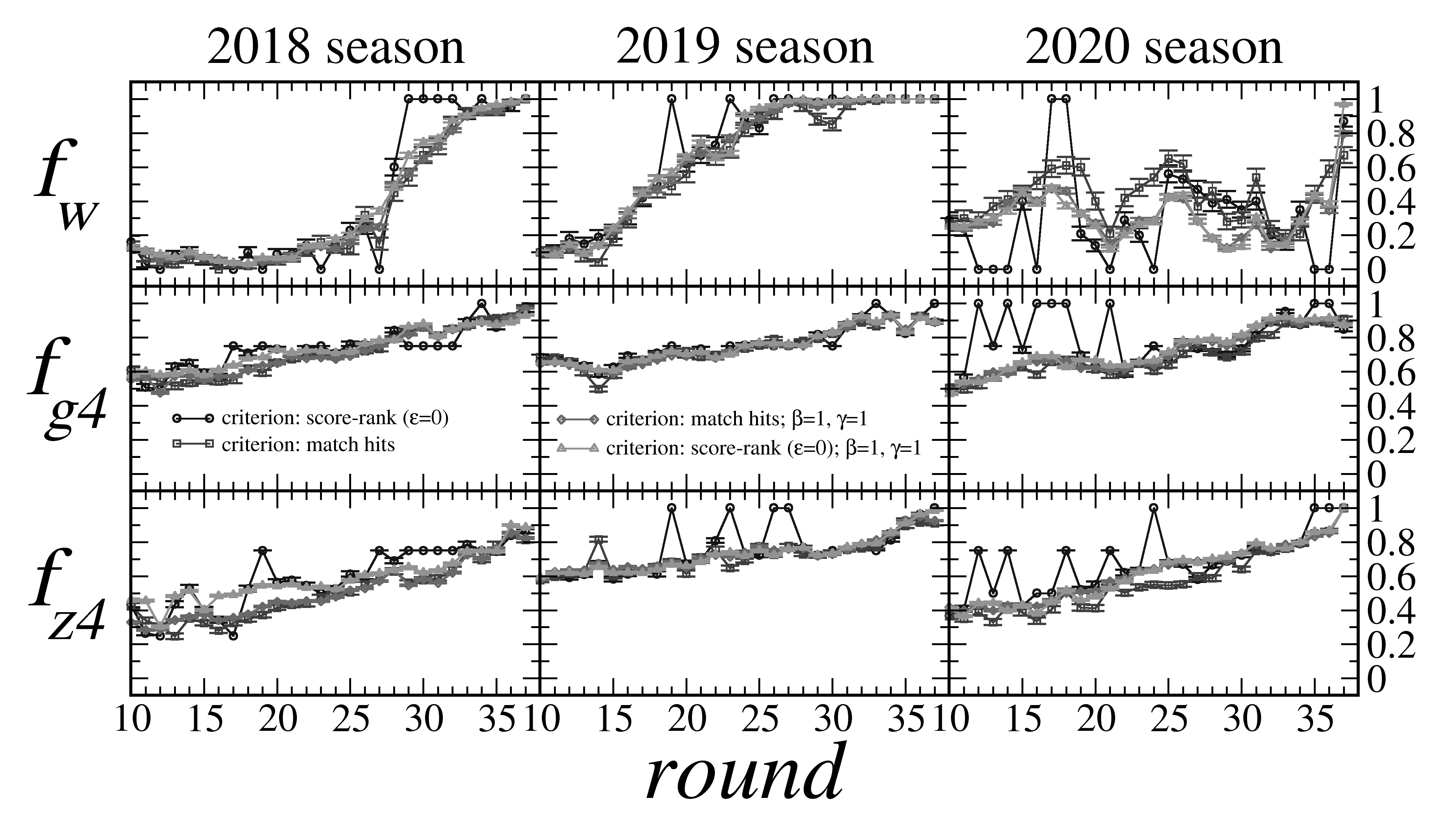}
\end{center}
\caption{Evolution of predictive parameters $\protect\mu _{w}$, $\protect\mu%
_{g4}$ and $\protect\mu _{z4}$ for the 2018, 2019, and 2020 seasons. We used 
$N_{run}=100$ to optimize and to predict the parameters. All situations lead
to good numbers for the predictions with a more oscillating evolution for
the score-ranking metric when using the optimization of the three
coeficients simultaneously. When considering the matching hits metric or by
fixing $\protect\beta =\protect\gamma =1$ the oscillation is reduced.}
\label{Fig:1st_2nd_model_2017-2019_comparison}
\end{figure*}

In this work, we are not only concerned with the champion of the league.
Instead, our goal is to compare the league champion, the top four teams
(usually denoted as $g4$) since their positions in the ranking table qualify
them to the stage groups of the continental cup known as Conmebol
Libertadores (Spanish and Portuguese for liberators) cup \footnote{%
http://www.conmebol.com/pt-br/torneos/conmebol-libertadores} (in the case of
South American clubs), and the last four teams (usually denoted as $z4$)
once they are relegated to a lower league division (Brazilian Championship B
Series) during the next season. For that, we define the following variables:

\begin{equation}
\begin{array}{rrrrrrr}
f_{w} & = & \frac{1}{N_{run}}\sum\limits_{i=1}^{N_{run}}\theta _{i} &  & 
f_{g4} & = & \frac{1}{4N_{run}}\sum\limits_{i=1}^{N_{run}}\iota _{i} \\ 
&  &  &  &  &  &  \\ 
&  & f_{z4} & = & \frac{1}{4N_{run}}\sum\limits_{i=1}^{N_{run}}\kappa _{i} & 
& 
\end{array}
\label{Eq:parameters}
\end{equation}%
where $\theta $ is equal to $1$ if the simulation got the champion right and 
$0$ otherwise, $\iota $ and $\kappa $ can assume integer values in the range 
$[0,4]$ depending on the number of teams that the simulations successfully
hit in the $g4$and $z4$ regions, respectively, of the ranking table.

The evolution of predictive parameters $f_{w}$, $f_{g4}$, and $f_{z4}$ for
the 2018, 2019, and 2020 seasons is shown in Fig. \ref%
{Fig:1st_2nd_model_2017-2019_comparison}. The black curve (circle) shows the
parameters obtained with score-ranking metric, dark gray (square)
corresponds to those obtained with matching hits criterion. Following, gray
(diamond) and light-gray (triangle) correspond to matching hits and
score-ranking metrics, respectively, by fixing $\beta =\gamma =1$, and only
the market value influence was optimized in this case. We can observe that
score-ranking metric with three optimized parameters presents a more
oscillating evolution. All situations are good in predicting $g4$, $z4$, and
the champion after a certain reasonable number of rounds. It is important to
mention that in the 2020 season, the prediction of the champion was more
complicated. Soccer Power Index (SPI) from the project FiveThirtyEight
(https://projects. vethirtyeight.com/soccerpredictions/brasileirao/ (2020))
corroborates this observation, but as we present below, our model was
capable of obtaining predictions as good or better as those of SPI. Thus,
let us summarize them in numbers!

\begin{table}[tbp] \centering%
\begin{tabular}{@{}ccccccccccccc}
\hline\hline
Season & \multicolumn{4}{c}{2017} & \multicolumn{4}{c}{2018} & 
\multicolumn{4}{c}{2019} \\ \hline\hline
$k$ & $25$ & $30$ & $35$ &  & $25$ & $30$ & $35$ &  & $25$ & $30$ & $35$ & 
\\ 
\multicolumn{1}{l}{score-ranking metric} & $1.00$ & $1.00$ & $1.00$ &  & $%
0.23$ & $1.00$ & $0.93$ &  & $0.83$ & $1.00$ & $1.00$ &  \\ 
\multicolumn{1}{l}{score-ranking metric (*)} & $0.87$ & $0.76$ & $1.00$ &  & 
$0.20$ & $0.74$ & $0.96$ &  & $0.94$ & $0.99$ & $1.00$ &  \\ 
\multicolumn{1}{l}{matching hits metric} & $0.67$ & $0.54$ & $1.00$ &  & $%
0.12$ & $0.67$ & $0.93$ &  & $0.89$ & $0.85$ & $1.00$ &  \\ 
\multicolumn{1}{l}{matching hits metric (*)} & $0.75$ & $0.68$ & $1.00$ &  & 
$0.16$ & $0.65$ & $0.92$ &  & $0.88$ & $0.97$ & $1.00$ &  \\ 
\multicolumn{1}{l}{SPI} & $0.78$ & $0.74$ & $0.99$ &  & $0.26$ & $0.77$ & $%
0.97$ &  & $0.94$ & $0.97$ & $1.00$ &  \\ \hline\hline
\end{tabular}
\caption{Hit estimates of the proposed model in relation to
the champion of the Brazilian Championship A Series in 2017, 2018, and 2019
seasons. Here (*) denotes the case where we fixed $\beta =1.0$, $\gamma =1.0$}%
\label{Table:final_estimates}%
\end{table}%

The results obtained for the champion of 2017, 2018 and 2019 seasons are
shown in Table \ref{Table:final_estimates}. For the sake of simplicity, the
results obtained for $g4$ and $z4$ ranking table are not shown in this
table. It is important to mention that such championships were well behaved
when compared to the 2020 season.

\begin{table}[tbp] \centering%
\begin{tabular}{cccccccccccc}
\hline\hline
$k$ & $25$ & $26$ & $27$ & $28$ & $29$ & $30$ & $31$ & $32$ & $33$ & $34$ & $%
35$ \\ \hline\hline
\multicolumn{1}{l}{score-ranking metric} & $0.56$ & 0.53 & 0.47 & 0.39 & 0.41
& $0.35$ & 0.40 & 0.20 & 0.15 & 0.35 & $0.00$ \\ 
\multicolumn{1}{l}{score-ranking metric (*)} & $0.42$ & 0.43 & 0.28 & 0.18 & 
0.11 & $0.13$ & 0.30 & 0.14 & 0.14 & 0.28 & $0.43$ \\ 
\multicolumn{1}{l}{\textbf{matching hits metric}} & $\mathbf{0.65}$ & 
\textbf{0.62} & \textbf{0.37} & \textbf{0.46} & \textbf{0.28} & $\mathbf{0.29%
}$ & \textbf{0.54} & \textbf{0.24} & \textbf{0.19} & \textbf{0.21} & $%
\mathbf{0.44}$ \\ 
\multicolumn{1}{l}{matching hits metric (*)} & $0.42$ & 0.41 & 0.28 & 0.18 & 
0.12 & $0.18$ & 0.26 & 0.12 & 0.17 & 0.27 & $0.42$ \\ 
\multicolumn{1}{l}{\textbf{SPI}} & $\mathbf{0.34}$ & \textbf{0.43} & \textbf{%
0.43} & \textbf{0.28} & \textbf{--} & $\mathbf{0.16}$ & \textbf{--} & 
\textbf{0.23} & \textbf{0.15} & \textbf{--} & $\mathbf{0.25}$ \\ \hline\hline
\end{tabular}%
\caption{Hit estimates of the proposed model in relation to
the champion of the Brazilian Championship A Series in 2020 season -- during
the pandemic. The bold lines stress our best metrics predictions compared 
with SPI predictions. We did not find predictions for the rounds 29, 31, and 34 made available by SPI. Here (*) denotes the case where we fixed $\beta =1.0$, $\gamma =1.0$}%
\label{Table:pandemic2020}%
\end{table}%

Our results also present strong correlations with those of SPI, as can be
seen in the last line of Table \ref{Table:final_estimates} showing that our
predictions are as good as the SPI ones (https://projects.
vethirtyeight.com/soccerpredictions/brasileirao/ (2020)). The predictions
are very similar, which is not exactly a surprise.

Finally, we would like to look into the numbers of a sui generis season:
2020. This season was held during the pandemic, which led to the absence of
fans in the stadium, and the champion (Flamengo) was known only at the end
of the last round, which dramatically affects the prediction. To get an
idea, Flamengo finished the championship with just 71 points, one point
ahead of Internacional, the vice-champion. Therefore, a single goal against
Corinthians in the last round would give the cup to Internacional, changing
the championship's history. As observed in Table \ref{Table:pandemic2020},
one of our versions is always better than the SPI predictions. In order to
perform a fair comparison, we choose our best case (matching hits metric)
that is better in all comparisons with SPI, except by the prediction of the
27th round where one has a dead heat.

\section{Summaries, conclusions, and discussions}

\label{Sec:Conclusions}

Our algorithm was capable of determining essential statistics of the real
Brazilian Championship A Series. Our predictions are compatible with an
alternative and commercial way to perform predictions (SPI) since, as shown,
both predictions are strongly correlated to each other. On the other hand,
our method presented more reliable estimates in a more complicated season:
2020. Our additional idea is to take a different path from SPI, making
explicit all of our ingredients. The model uses many concepts related to the
teams' scoring process during a championship and the market values of the
teams.

In addition, the method can be easily applied in any championship to obtain
good probabilities of hitting the champion. Estimates for $g4$ and $z4$ show
notable agreement between the metrics considered when the predictions are
performed from the 26th round on. In fact, from this round on, one obtains
something between 80\%--100\% of probability to point out the exact $g4$ or $%
z4$ (see Fig. \ref{Fig:1st_2nd_model_2017-2019_comparison}). We believe that
our predictive method can be changed to contemplate other championships
based on the double-round robin system scheme. Other authors \cite%
{Baboota2018} explored machine learning for the premier league in seasons
2014 and 2015 with good predictions, and the topic seems to be extremely
promising for future works by including other sports.

\textbf{Acknowledgements}

The authors thank CNPq for financial support under grant numbers
311236/2018-9 and 424052/2018-0.

\bigskip

\end{document}